# The origin of atomic displacements in HAADF images of the tilted specimen

J. Cui, Y. Yao*, Y. G. Wang, X. Shen, R. C. Yu

Beijing National Laboratory of Condensed Matter Physics,

Institute of Physics, Chinese Academy of Sciences,

Beijing 100190, China

*Corresponding author, email: yaoyuan@iphy.ac.cn

Abstract

The effects of the tilt of the crystallographic orientation with respect to an incident electron probe on high-angle annular dark field (HAADF) imaging in aberration-corrected scanning transmission electron microscopy (STEM) have been investigated with experiments and simulations. A small specimen tilt can lead to unequal deviations of different atom species in the HAADF image and result in further relative displacement between anions and cations. Simulated HAADF images also confirm that the crystal tilt causes an artifact in atom polarization as well. The effect is derived from the scattering ability of different atoms.

Introduction

The high-angle annular dark field (HAADF) images of the Cs-corrected scanning transmission electron microscopy (STEM) plays an important role in characterization of the crystal structure with picometer-precision. The HAADF image is thought to intuitively interpret atomic column positions over a relative wider range of specimen thickness and the defocus condition because of its incoherent contrast avoiding the reverse from the phase contrast that inheres in conventional transmission electron microscopy (TEM). In addition, the HAADF image contrast depends strongly on the atomic number Z of the scattering atoms in a simple $Z^n$ form (n=1.6-1.9)[1] except for some special conditions[2], which makes the composition identification easy through intensity patterns in the image. Recently some literatures[3-6] described a direct estimate of the atom displacement by characterizing the bright spot displacement in the HAADF image under the hypothesis that those bright spots correspond to the actual positions of the atomic columns. Spurgeon et al. directly measured the induced ferroelectric polarization from the relative position deviation of the heavy atoms in the $SrTiO_3$ layers and revealed that the built-in asymmetric potential across the $LaCrO_3/SrTiO_3$ interfaces is sufficient to induce a sizable polarization, on the order of 40-70 μC cm$^{-2}$, in good agreement with *ab* initio calculations.[3] Li et al. directly observed the ferroelectric flux-closure vortex domains at the $BiFeO_3/TbScO_3$ hetero-interfaces using the STEM-HAADF. They found the Fe cation displacing from the center of the unit cell formed by its four Bi neighbors in the HAADF image and showed that a size effect on the spontaneous flux-closure domains could induce the formation of smoothly and continuously rotated polarization patterns in ultrathin ferroelectrics.[4] However the assumption that bright spots present the projection of atomic columns perpendicular to the image plane should be examined carefully because contrast patterns in the HAADF image could be influenced by many experimental factors, such as specimen tilt[7-10], specimen thickness[8,9], convergence semi-angle of the electron beam[8], collection angles of the detector[8], etc. So et al.[8] reported that a small sample tilt causes an artifact deviation in the distance between Sr and Ti atoms in the perovskite

SrTiO$_3$. The displacement of each atomic chains in HAADF image differs when the specimen inclines slightly away its [001] axis, resulting in the noncentrosymmetrical [001] projection of the crystal. The displacement is thought from the electron channeling to adjacent atomic columns and a large convergence angle is suggested to enable the precise measurement of the positions if the Cs-corrected STEM is utilized. Wang et al. Characterized the SrTiO3 (001) interfaced with an epitaxial La$_{2/3}$Sr$_{1/3}$MnO$_3$ film and indicated that the sample bending could cause a considerable displacement of the atoms in HAADF images.[10] Some researches explored the tilt effect on annular bright-field (ABF) images, which is severer than the HAADF image.[11, 12] Zhou et al. studied the effect of tilted electron beam relative to the crystal zone axis on the determination of atom positions in the ABF image. They found that in ABF images, aligning the specimen as close as possible to the targeted zone axis with assistance of the techniques like position averaged convergent beam electron diffraction(PACBED) can improve the accuracy of atomic column position determination; selection of larger convergence semi-angles and corresponding larger collection semi-angles can make smaller errors in determining the relative positions or angles between different atomic columns on average, but not for all thickness situations; the heavier the element is, the smaller the overall deviation will be with a rapid variation occurring within certain thickness.[11] Moreover, Gao et al. investigated the similar effects of the crystal tilt on the precision measurement of ABF images and found that the tilt induced artificial displacements depend on the atom species, defocus, thickness of specimen and convergence angle.[12] They also indicated the abnormal deviation of the atoms in the HAADF simulation images.

In this paper, we characterize such tilt effects on the measurement of atomic chain positions in InAs/GaSb superlattice with a large convergence angle in the Cs-corrected transmission electron microscope. The displacement of cations relative to the mass center of four neighboring anions is examined in the centrosymmetry [100] projection. With the assistance of simulation, the artificial displacement between anions and cations is ascribed to the different scattering contributions of the atoms involved along the incident electron beam.

Methods

The investigated InAs/GaSb superlattice was grown on (100) GaSb substrates by molecular beam epitaxy. The modulated superlattices consist of 8 monolayers of GaSb and InAs, respectively, alternating along the growth direction. Cross-sectional samples were prepared by FEI Helios NanoLab 600i DualBeam FIB/SEM for TEM observations along the [100] zone axis. Scanning transmission electron microscopy (STEM) experiments were performed on an aberration-corrected JEOL JEM-ARM200F microscope operated at 200 kV acceleration voltage. The experimental convergence semi-angle was about 22 mrad and the corresponding collection angle range was 70-370 mrad. To obtain the atomic column positions, we used peak pair algorithm (PPA) developed by Galindo et al.[13] This method is implemented as a commercial plug-in (HREM Research, Inc.) to the software Digital Micrograph marketed by Gatan, Inc.

The simulated HAADF-STEM images presented here were carried out using the multislice algorithm accelerated by CPU/GPU hybrid computing.[2] The size of the unit cells for the simulation were $6.096 \times 6.096 \times 6.096$ Å$^3$, $6.058 \times 6.058 \times 6.058$ Å$^3$ and $6.096 \times 30.292 \times 6.096$ Å$^3$ for the pure GaSb, pure InAs and the artificial InAs/GaSb superlattice. STEM images were simulated with 0.02 Å sampling in real space but the scanning step of the electron beam is 0.1 Å and 0.02 Å for image and line-scanning simulation, respectively. Not the electron beam, but the specimen was tilted 0 °, ±0.2 °, ±0.5 °, ±1 ° for all the configurations of simulation and the tilt axis was [010] direction, which lay at the superlattice interface and was perpendicular to the growth [001] direction. All images were simulated with the following parameters: acceleration voltage was 200 kV, spherical aberration of *Cs* was 0 mm, defocus of *Δf* was 0, convergence semi-angle was 22 mrad, collection angle was between 60 and 200 mrad.

Experiments

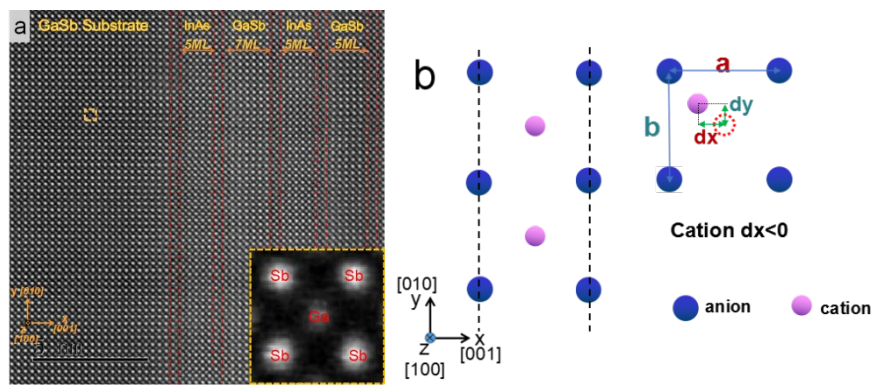

Fig. 1. (a) The filtered HAADF image of InAs/GaSb superlattice along [100] zone axis. The inset is a magnified image showing the chosen unit cell of atoms. The electron beam scanned from left to right for all the HAADF images in this paper. (b) The definition of *a*, *b* and the displacement of cation relative to the mass center of four neighboring anions.

Fig. 1a displays the HAADF image of a InAs/GaSb superlattice including the GaSb substrate, filtered based on the Kilaas's method[14]. It is easy to distinguish the atom species from the Z contrast of the spots in the substrate, where the most bright ones should be assigned as Sb (Z=51) and the others are Ga (Z=31), then In and As could be deduced in the superlattice structure by contrast or by the cation/anion sequences as well. Considering the binding energy of GaAs (3.35eV), InAs (3.17eV), GaSb (2.88eV), InSb (2.27eV),[15] the atoms may diffuse and mix in the interface of InAs/GaSb layers due to the favorite Ga-As structure with a higher binding energy,[16] which results in mutual contrast along the interfaces, as marked by the red dashed lines. To analyze the deviations, the basic unit cell of atoms was chosen as displayed in the inset of Fig. 1a, where the anion atoms are in the corners and the cation atom locates in the center of the anions square for a perfect [100] projection. For the convenience, the positions of the atoms were denoted as *x-y* coordination and the *x* and *y* axes were defined along the [001] and [010] directions, respectively. The distance between anions along *x* and *y* directions were denoted by *a*, *b*, respectively, as marked in Fig. 1b. The interatomic distances between each spots were calculated as well. Helped with PPA, the positions and intensities of each cation and anion were recognized and averaged along [010] direction to reveal the subtle position deviation among the atoms. Using a homemade DM script, the relative shifts *dx* and *dy* of cation relative to the mass center of four neighboring anions could be acquired. The positive and negative of displacements are defined in Fig. 1b. The intensity profile of a line scanning along the [001] growth direction was used roughly to draw the regions of the superlattice in the following paragraph.

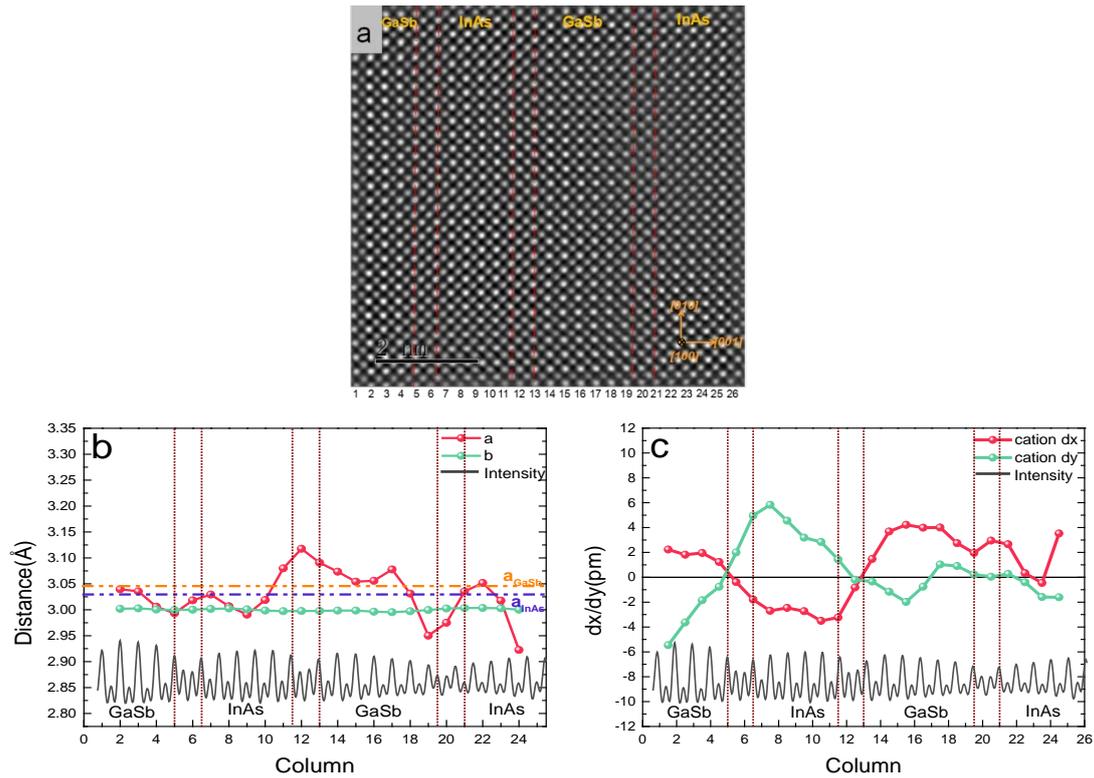

Fig. 2. (a) The filtered HAADF image of few InAs/GaSb periods adjacent to the substrate, where the corresponding column numbers are marked below. (b) Averaged *a* and *b* of each column in (a) with the intensity profile (below). The unit size of pure GaSb ($a_{GaSb}$ =3.05 Å) and InAs ($a_{InAs}$ =3.03 Å) are denoted by the dash-dotted lines, respectively. (c) The deviations of cations with the intensity profile (below). The vertical dotted lines indicate the superlattice interfaces.

Fig. 2a is the filtered HAADF STEM image of few superlattice periods near the substrate. Fig. 2b depicts the *a* and *b* of each column along the growth direction, which were averaged in the corresponding column to mediate the subtle drift during the electron beam scanning. It looks that *a* is greater than the unit cell constant of pure GaSb (3.05 Å) in GaSb layers, but less than that of pure InAs (3.03 Å) in the InAs layers, both of which are shown by the dash-dotted lines. Meanwhile, the *b* remains nearly constant among the whole superlattice but is less than the size of pure InAs, which implies that there may be some systemic errors in the image calibration. Few changes of *b* reflect the perfect in-plane match of the superlattice but such match induces the appearance of variation in *a* due to the strain from the intrinsic lattice mismatch between GaSb and InAs layers. The relative shifts *dx* and *dy* of cation chains relative to the mass center of four neighboring anion chains are plotted in Fig. 2c. Obviously, positive displacements *dx* are revealed in GaSb region, while the negative displacements are in InAs region. Similar behavior can be found in the *dy*. It seems that there is a local polarization within each layer of the superlattice but opposite between different layers.

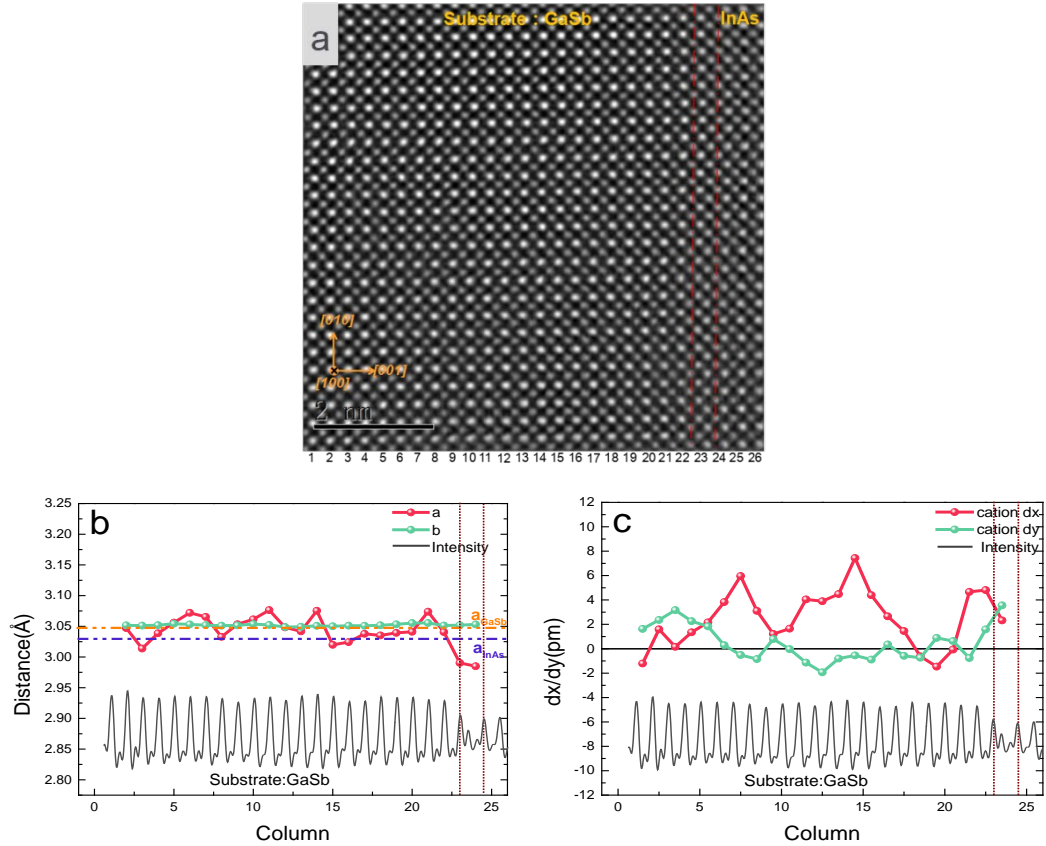

Fig. 3. (a) The filtered HAADF image of the GaSb substrate and the first InAs layer in InAs/GaSb superlattice, where the corresponding column numbers are marked below. (b) Averaged *a* and *b* of each column in (a) with the intensity profile (below). The unit sizes of pure GaSb ($a_{GaSb}$ =3.05 Å) and InAs ($a_{InAs}$ =3.03 Å) are denoted by the dash-dotted lines, respectively. (c) The deviations of cations with the intensity profile (below). The vertical dotted lines indicate the superlattice interface.

Fig. 3a is the filtered HAADF STEM image including the GaSb substrate and the first InAs layer of the superlattice. As seen in Fig. 3b, in GaSb region, the value of *a* fluctuates around *b*. Fig. 3c shows that the displacements *dx* of cation chains are generally positive in the GaSb substrate. However the substrate region should be uniform and non-polarized, which is inconsistent with the experimental data. Obviously, such displacements of atoms need further examination into the imaging mechanism of HAADF images.

As reported in the literature, the crystal zone axis tilts away from the electron beam may cause the displacements of atoms in STEM images. Thus HAADF images of specimen were intentionally acquired with and without tilt to verify the origin of the atom displacement. When the specimen was under crystal [100] zone axis, the average *dx* in the GaSb substrate is about 0.5 pm and *dy* is about -1.9 pm with an alternative "polarization" in the staggered superlattice, as shown in Fig. 4a. After the specimen was tilted 1° along *y* ([010]) direction, there are obvious enhancements of *dx* and *dy* in the substrate while a reversal *dx* in the superlattice where the *dy* is just a little larger, as plotted in Fig. 4b. It is evidential that the sample tilting could lead to the plausible "polarization" in the centrosymmetric zincblende structure.

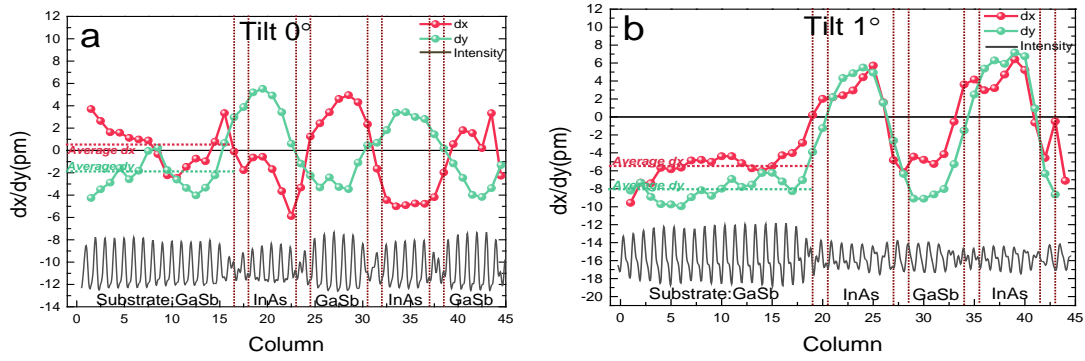

Fig. 4. The statistic results from HAADF images of deviations of cations relative to the mass center of four neighboring anions in no tilt (a) and 1° tilt (b) specimen. The average *dx* and *dy* in the substrate GaSb region are denoted by green and red dash lines, respectively.

Simulation

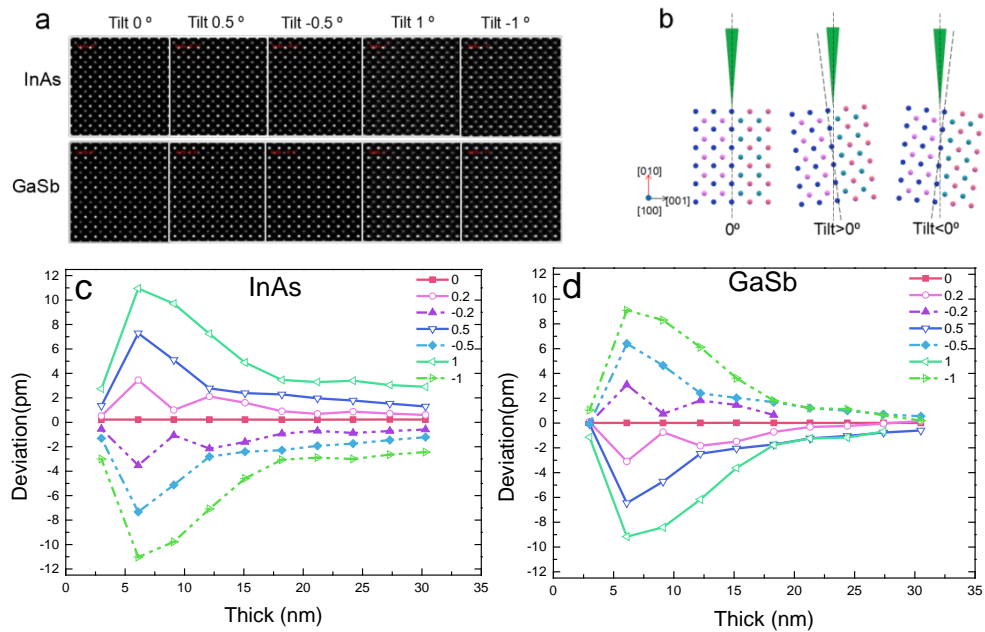

Fig. 5. (a) The simulated HAADF images of InAs (100) single crystal and GaSb (100) single crystal with a thickness of 24.3 nm for tilt angle=0°, ±0.5° and ±1°, respectively. (b) The illustration of specimen tilt relative to the incident electron beam. Deviations of cation chains relative to the mass center of four neighboring anion chains in InAs(c) and GaSb(b) with different specimen thicknesses and tilt angles, respectively.

To reveal the detail of tilting effect on the accurate measurement of atom positions in a HAADF image, firstly, a series of HAADF images of GaSb and InAs single crystal tilted away from the incident electron beam and with the thicknesses up to 50 unit cells were simulated without consideration of Debye-Waller factor, which are shown in Fig. 5. In Fig. 5a, it is not easy to discern the changes in the spots at a small tilt angle. For a large tilt angle (±1°), the little elongation of the bright spots is more striking both in simulated InAs and GaSb, which reflects that the large tilt angle exerts a greater influence on the contrast of heavier atoms in HAADF images. Fig. 5b and c plot statistic cation displacements *dx* with different tilt angles and specimen thicknesses of InAs and GaSb, displaying that *dx* increases within the thickness of 6nm, but gradually decreases for the thicker specimen. A larger tilt causes a larger deviation *dx*. Moreover, the cations deviate inversely under the same tilt angle in the InAs and GaSb single crystals.

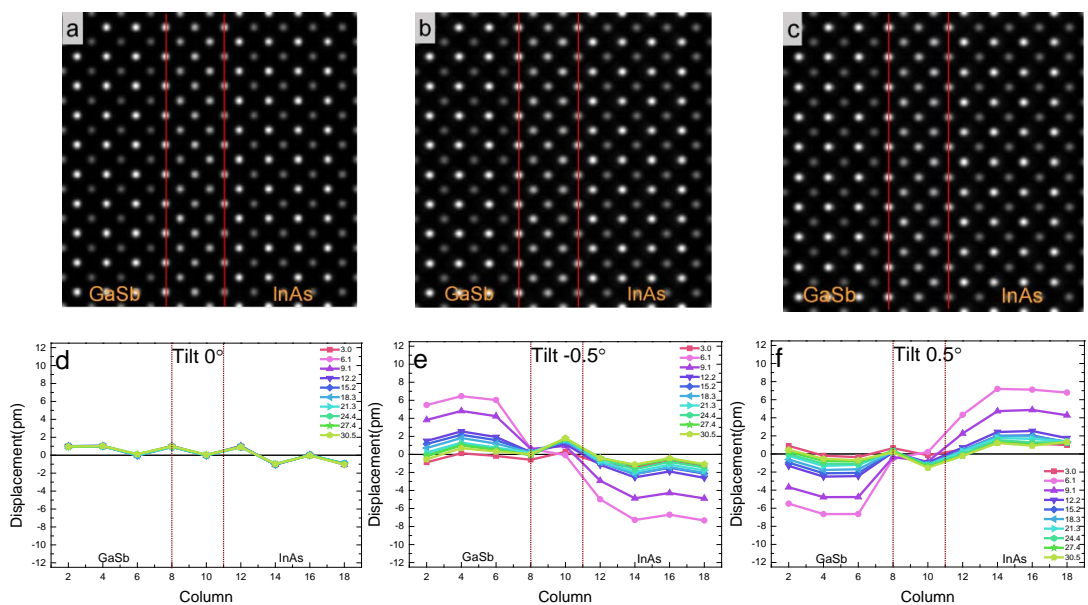

Figure. 6. The simulated HAADF images of an artificial InAs/GaSb superlattice with a specimen thickness of 12.2 nm for tilt angles = 0° (a), -0.5° (b) and 0.5° (c), respectively. The statistic deviation of cation chains relative to the mass center of four neighboring anion chains in the InAs/GaSb superlattice with different specimen thicknesses tilt at angle=0° (d), -0.5° (e) and 0.5° (f), respectively. The interface is $In_{0.5}Ga_{0.5}As_{0.5}Sb_{0.5}$ for the convenience.

Fig. 6 shows the simulated HAADF images of InAs/GaSb superlattice by multislice method with the frozen phonon model. For the convenience, the interface of InAs/GaSb is two unit cells wide with an equal mixed occupancy. The simulated InAs/GaSb superlattice is tilted at 0°, -0.5° and 0.5° with the thickness from 3.0 nm to 30.5 nm. When the specimen tilt angle is 0°, cation displacement *dx* oscillates around zero about 0.5 picometer both in GaSb and InAs regions in any thickness. If the specimen tilt angle is 0.5°, *dx* is negative in GaSb region, gradually changing in the interfaces region, and becomes positive in InAs region, showing an inversed "polarization". In addition, such a divergence rises initially but then falls with increasing thickness. These simulated images reproduce the experimental results. Therefore, the "polarization" may be an artifact from the specimen tilt.

Discussion

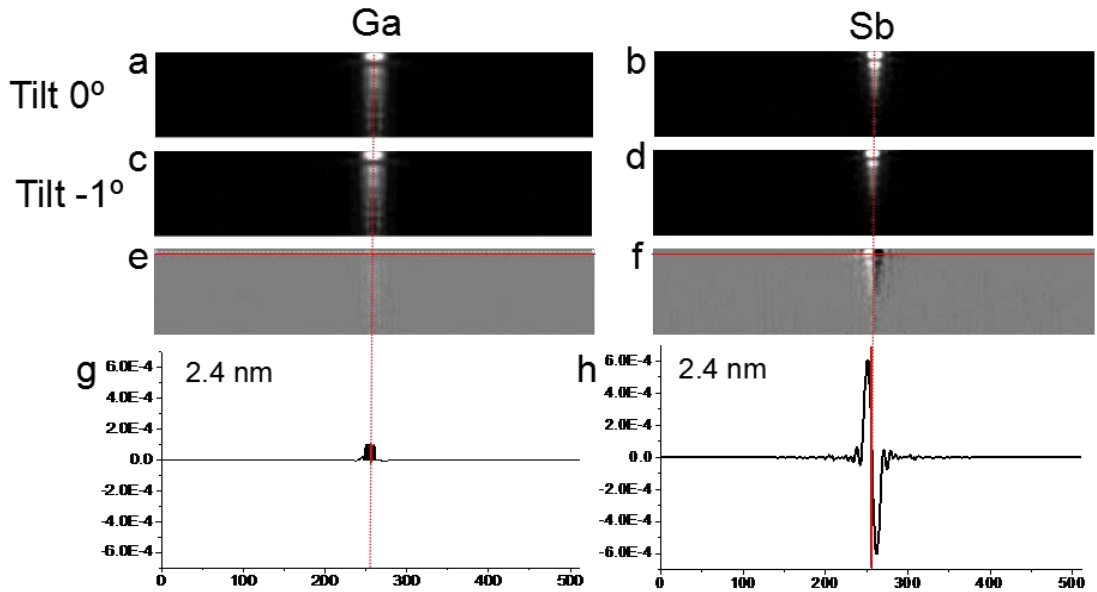

Figure. 7. The simulated intensity of electron wave propagating through the InAs/GaSb superlattice when the electron probe focuses on the Ga (left column) and Sb (right column) chains. The probe intensity is shown in grayscale at different depths of the specimen and the cross section is along [001] direction. (a) and (b) without tilt, while (c) and (d) tilt -1°. (e) The intensity difference of (a) and (c). (f) The intensity difference of (b) and (d). (g) and (h) are the profiles of intensity difference at 2.4nm depth in (e) and (f), respectively.

So et al. attributed similar phenomenon in the $SrTiO_3$ sample to the focus depth of the probe beam, which extends the incident electrons into adjacent atomic columns and induces a complex contrast distribution when the [001] orientation is tilted away from the incident beam. They also suggested that a larger convergence angle of the electron beam with a short focus depth should enable an accurate identification of the atomic position because of avoiding the crosstalk between different atomic columns.[8] But the experimental and simulated HAADF images still display the evident displacement of the cation relative to the center of the four anions even for the 22 mrad convergence beam here. In order to explore the reason of that displacement, the intensity of the electron wave propagating in the crystal was extracted from the simulation results to elucidate the scattering process of each atomic column. Fig. 7 shows the simulated electron intensity transmitting along the Ga and Sb chains in the InAs/GaSb superlattice with and without sample tilt. The transmission electrons move more densely in the Sb chains than in the Ga chains, reflecting that the channel effect is stronger for the heavier Sb atoms due to their larger Z. The intensity difference between the tilted and non-tilted specimens in Fig. 7e and f reveals that the orientation change induces tiny variation for Ga chains but alters the intensity severely for Sb chains. This means that the electrons should transport along the Sb chains for a longer distance even if the incident beam deviates the exact direction of the atomic columns, as profiled in Fig. 7 g and h, respectively. The channel effect in the heavier Sb chains is a focus-like behavior and prolongs the propagation length of electrons around the Sb chains. In other words, the "effective" length or the number of atoms involved in scattering along the chains is not equal for Ga and Sb; the heavier the atom is, the more atoms contribute to the scattering process.

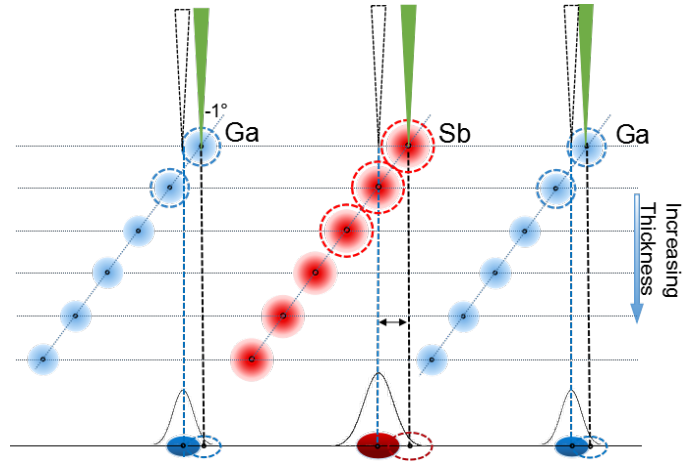

Figure. 8. Schematic of different atomic scattering contributions and their effects on the displacements of atomic columns projection.

The different effects on the displacements of Ga and Sb atomic columns under same tilt angle are attributed to their different atomic scattering properties and dynamical effects, which is depicted schematically in Fig. 8 where the [100] axis is tilted -1° from the incident beam. The red circle denotes Sb atom and blue circle denotes Ga atom. Big atomic radius represents strong atomic scattering potential and dotted circled atoms represent roughly the involved atoms which play the main roles in scattering. Heavier Sb atom leads to a deeper scattering range of Sb chain even if the specimen is tilted. Therefore, the maximum of the scattering intensity of the Sb in the HAADF image should be further away from the projection of the first layer atoms. So if the maximum of Ga spot is set as the reference, the maximum of Sb spot will move toward the left when the specimen is tilted negatively. Consequently, for InAs sample, the movements of the maximum of the atoms are reversal in same tilt angle because the scattering properties of the cation and anion are reversal: the lighter As atom seems to shift to its right In neighbor in the HAADF image. The difference in scattering ability between the atomic columns is an essential influence for the relative displacement of the atoms in HAADF images if sample is tilted.
The non-equivalence between the displacements of heavy and light atoms in the tilted specimen is the result of the intrinsic quality of the electron-specimen interaction in the transmission electron microscopy, different from other instrument parameters. The artificial displacement occurring strongly in the first few nanometers depth of the specimen infers that the influence of the channeling effect should be the main source because the de-channeling does not appear at this depth. Larger convergence probe could weaken this effect by avoiding the crosstalk between the adjacent atomic columns but could not remove it completely. Unfortunately, considering the real operation conditions, it is not easy to ensure that the axis of specimen points to the incident beam exactly when taking the HAADF images. The HAADF images themselves cannot supply enough evidence to judge whether the tilt exists or not because both experimental and simulation images show the clear atomic resolution pictures when the samples are tilted slightly. There is always a small bending of the thin TEM specimen, especially for the samples with interfaces, defects or atom diffusion and other inherent strain; the imaging field is not the exact place where the probe beam locates for the specimen orientation adjustment; a tiny error of alignment with Kikuchi pattern is not discernible to the eyes of the operator. So sometimes a

"good" HAADF image taken with an imperceptible specimen tilt, which cannot be identified before or after the acquirement, will exhibit a measurable artificial atomic shift in the following image processing, as shown in Fig. 4a where the image was acquired in a seemingly correct alignment but an evidential *dx* or *dy* still emerges. Even for the real polarized structure, the tilt effect can lead to errors in the displacement measurement. Thus further techniques should be developed to improve the specimen alignment for STEM imaging.[10, 11, 17]

Conclusions

In summary, the influence of specimen tilt on the InAs/GaSb superlattice atomic displacements in HAADF STEM was investigated. The experimental data indicate that the specimen tilt can lead to inequivalent deviation of different atomic species and result in the relative displacements between anions and cations. Simulation images disclose that this inequivalent shift arises from the interaction of the electrons with the specimen, reflecting the different scattering abilities of different atoms. The artificial atomic displacement caused by the specimen tilt should be taken into consideration when analyzing the atomic resolution HAADF images.

Acknowledgments

This work was supported by the Ministry of Science and Technology of the People's Republic of China (No. 2016YFA0202500), the State Key Development Program for Basic Research of China (No. 2013CB932904) and the National Natural Science Foundation of China (Nos. 11374343 and 11574376).